\documentclass[10pt,conference]{IEEEtran}
\IEEEoverridecommandlockouts
\usepackage{cite}
\usepackage{amsmath,amssymb,amsfonts}
\usepackage{algorithmic}
\usepackage{graphicx}
\usepackage{textcomp}
\usepackage{xcolor}
\usepackage{hyperref}
\usepackage{xurl} 
\usepackage{array}

\hypersetup{
  colorlinks=true,
  linkcolor=IEEEblue,
  citecolor=IEEEblue,
  urlcolor=IEEEblue
}
\definecolor{IEEEblue}{HTML}{00629B}   

\renewcommand{\thesection}{\arabic{section}}
\renewcommand{\thesectiondis}{\arabic{section}}
\renewcommand{\thesubsection}{\thesection.\arabic{subsection}}
\renewcommand{\thesubsectiondis}{\thesection.\arabic{subsection}}

\makeatletter
\newcommand{\linebreakand}{%
  \end{@IEEEauthorhalign}
  \hfill\mbox{}\par
  \mbox{}\hfill\begin{@IEEEauthorhalign}
}
\makeatother

\usepackage{fancyhdr}

\AtBeginDocument{%
  \fancypagestyle{firstpage}{%
    \fancyhf{}%
    \fancyfoot[C]{\footnotesize Accepted for the \textit{2nd IEEE/ACM International Conference on AI-powered Software (AIware 2025), November 19-20, 2025, Seoul, South Korea.}}%
  }%
}

\begin{document}

\title{The Future of Generative AI in Software Engineering: A Vision from Industry and Academia in the European GENIUS Project
\thanks{This work was carried out within the ITEA 4 project GENIUS, as part of the ITEA programme, the Eureka Cluster on software innovation.
This work was funded by the German Federal Ministry of Research, Technology and Space (BMFTR) under grant numbers 16IS24069A, 16IS24069B, 16IS24069E, 16IS24069G, and 16IS24069H, the Austrian Research Promotion Agency (FFG) under grant numbers 931318 and 921454, InnovateUK under grant number 600642, and by TÜBITAK under grant number 9240014.}}


\author{
\IEEEauthorblockN{\ Robin Gröpler\textsuperscript{*}\thanks{*Corresponding author: Robin Gröpler, \href{mailto:robin.groepler@ifak.eu}{robin.groepler@ifak.eu}}}
\IEEEauthorblockA{\textit{ifak e.V.}\\
Magdeburg, Germany}
\and
\IEEEauthorblockN{Steffen Klepke}
\IEEEauthorblockA{\textit{Siemens AG}\\
Munich, Germany}
\and
\IEEEauthorblockN{Jack Johns}
\IEEEauthorblockA{\textit{BT Group}\\
Ipswich, UK}
\and
\IEEEauthorblockN{Andreas Dreschinski}
\IEEEauthorblockA{\textit{Akkodis Germany Solutions GmbH}\\
Sindelfingen, Germany}
\and
\IEEEauthorblockN{Klaus Schmid}
\IEEEauthorblockA{\textit{University of Hildesheim}\\
Hildesheim, Germany}
\linebreakand
\IEEEauthorblockN{Benedikt Dornauer}
\IEEEauthorblockA{\textit{University of Innsbruck}\\
Innsbruck, Austria}
\and
\IEEEauthorblockN{Eray Tüzün}
\IEEEauthorblockA{\textit{Bilkent University}\\
Ankara, Türkiye}
\and
\IEEEauthorblockN{Joost Noppen}
\IEEEauthorblockA{\textit{BT Group}\\
Ipswich, UK}
\and
\IEEEauthorblockN{Mohammad Reza Mousavi}
\IEEEauthorblockA{\textit{King’s College London}\\
London, UK}
\and
\IEEEauthorblockN{Yongjian Tang}
\IEEEauthorblockA{\textit{Siemens AG}\\
Munich, Germany}
\linebreakand
\IEEEauthorblockN{Johannes Viehmann}
\IEEEauthorblockA{\textit{Fraunhofer FOKUS}\\
Berlin, Germany}
\and
\IEEEauthorblockN{Selin Şirin Aslangül}
\IEEEauthorblockA{\textit{Beko}\\
Istanbul, Türkiye}
\and
\IEEEauthorblockN{Beum Seuk Lee}
\IEEEauthorblockA{\textit{BT Group}\\
Ipswich, UK}
\and
\IEEEauthorblockN{Adam Ziolkowski}
\IEEEauthorblockA{\textit{BT Group}\\
Ipswich, UK}
\and
\IEEEauthorblockN{Eric Zie}
\IEEEauthorblockA{\textit{GoCodeGreen}\\
London, UK}
}

\maketitle


\begin{abstract}
Generative AI (GenAI) has recently emerged as a groundbreaking force in Software Engineering, capable of generating code, identifying bugs, recommending fixes, and supporting quality assurance. While its use in coding tasks shows considerable promise, applying GenAI across the entire Software Development Life Cycle (SDLC) has not yet been fully explored. Critical uncertainties in areas such as reliability, accountability, security, and data privacy demand deeper investigation and coordinated action. 
The GENIUS project, comprising over 30 European industrial and academic partners, aims to address these challenges by advancing AI integration across all SDLC phases. It focuses on GenAI’s potential, the development of innovative tools, and emerging research challenges, actively shaping the future of software engineering. This vision paper presents a shared perspective on the future of GenAI-driven software engineering, grounded in cross-sector dialogue as well as experiences and findings within the GENIUS consortium. 
The paper explores four central elements: (1) a structured overview of current challenges in GenAI adoption across the SDLC; (2) a forward-looking vision outlining key technological and methodological advances expected over the next five years; (3) anticipated shifts in the roles and required skill sets of software professionals; and (4) the contribution of GENIUS in realising this transformation through practical tools and industrial validation. This paper focuses on aligning technical innovation with business relevance. It aims to inform both research agendas and industrial strategies, providing a foundation for reliable, scalable, and industry-ready GenAI solutions for software engineering teams.
\end{abstract}


\thispagestyle{firstpage}

\section{Introduction}
Generative AI (GenAI) and, in particular, Large Language Models (LLMs) are advancing rapidly and have already shown their significant potential to transform software engineering practices. Tools such as GitHub Copilot \cite{copilot}, Amazon Q \cite{amazonq}, and Anthropic Claude Code \cite{claudecode} demonstrate productivity gains in coding, marking a paradigm shift in how software is developed \cite{jiang2024survey}. Yet, despite recent advances, the systematic integration of GenAI throughout the Software Development Lifecycle (SDLC), spanning requirements engineering, design, implementation, and testing, remains fragmented and insufficiently understood \cite{hou2024large, zhang2023survey, fan2023large}. 

Existing GenAI adoption remains largely confined to isolated phases, overlooking the opportunities and challenges of cross-phase integration, human–AI collaboration at scale, and deployment in real-world industrial ecosystems. Current LLM-based approaches still lack sufficient contextual grounding, domain adaptation, and reasoning capabilities to support complex engineering workflows \cite{jin2024llms}. These limitations highlight the need for a coherent vision that connects academic insight, technological innovation, and industrial experience into a unified roadmap for the future of GenAI-driven software engineering. Such a vision must bridge research and industrial practice, span the full SDLC, and foster large-scale collaboration across multiple domains \cite{nguyenduc2023generative}. 

This vision paper combines a forward-looking scientific perspective with practical insights from the GENIUS project \cite{itea-genius}, a large-scale European research initiative that aims to develop automated solutions and customised tools to enhance and bridge the various phases of the SDLC. GENIUS brings together leading industrial partners, applied research institutes, and universities to jointly advance GenAI-driven methods and evaluate their impact in industrial development environments. The ideas presented here have emerged from intensive research work and extensive discussions among the partners, combining academic and industrial perspectives grounded in emerging experiences from real-world industrial settings. The paper aims to articulate the evolving role of GenAI in software engineering and to examine its broader implications for processes, tools, and human roles across the SDLC \cite{prather2024hype}. 

In the following, we first identify in Section \ref{sec:challenges} the key challenges that currently constrain the reliable adoption of GenAI across the SDLC. Section \ref{sec:vision} envisions and discusses the potential future of GenAI-driven software engineering, outlining the capabilities, transformations, and evolving human skills required for sustainable and trustworthy GenAI ecosystems. Section \ref{sec:outlook} then connects these conceptual foundations to their practical realisation within the GENIUS project, where the envisioned approaches are operationalised and validated across diverse industrial domains.

\section{Challenges of GenAI in Software Engineering} \label{sec:challenges}

The current application of GenAI in software engineering faces various challenges spanning different stages of the SDLC as well as aspects of data, training, and evaluation of GenAI. This section provides an overview of the challenges from our perspective – based on our experience and findings. 

\subsection{Hallucinations, Limited Reasoning, and Challenges with Structured Output Generation} \label{sec:hallucinations}

LLMs are trained on large public datasets, including code repositories, that often include inconsistent, uncurated, or outdated information. As a result, LLMs are exposed to learn from flawed information. Additionally, their probabilistic nature can lead to confident but incorrect or unverifiable outputs, known as hallucinations \cite{kalai2025why}. For example, it is common for LLM-generated source code to use deprecated libraries or functions, potentially introducing security vulnerabilities, and creating increased technical debt leading to code bases which are difficult to maintain \cite{gao2024current}.

Additionally, LLMs have limited reasoning capabilities because they rely on statistical pattern matching rather than genuine contextual understanding, e.g., of source code bases. Their responses often replicate reasoning patterns seen in training data, rather than deriving conclusions through logical deduction \cite{chen2025deep}. Furthermore, studies found that current LLMs often lack sufficient awareness of the user's expertise and specific needs, providing responses that are either too complex or too simplistic for the user's skill level \cite{tie2024llms}. 

Reliable generation of highly structured outputs, such as programme code, configuration files, or domain-specific languages, remains challenging for LLMs. One common approach to address this issue involves guiding the model’s output using predefined grammatical rules -- known as Grammar-Constrained Decoding (GCD). For example, a specific grammar like a context-free grammar designed for a domain-specific language can be used to limit which words or symbols the model is allowed to generate. This helps to ensure that the output follows correct syntax, even if the model was not explicitly trained on that grammar. However, strictly enforcing such rules can interfere with the model’s natural generation process and may lead to low-quality results. Furthermore, these methods can be technically demanding to set up and may become slow or inefficient, especially when working with large or complex grammars \cite{netz2024using, park2024grammar}. 

The concern surrounding reliability causes further problems within the software testing phase. Due to the probabilistic outputs produced by AI models, slight changes in input data can lead to significant variations in results, complicating the establishment of reliable test oracles \cite{gao2024current}. 

\subsection{Limited Context Awareness and Poor Support for Domain-Specific Knowledge} \label{sec:context}

GenAI models often struggle to understand specific contexts, such as those of large requirements documents or large source code bases, leading to less precise or relevant suggestions. As a result, LLMs may not fully grasp given specific requirements, source code dependencies, or architectural goals of a software engineering project \cite{gao2024current}.  

Current LLMs show limited understanding of abstract software engineering principles, such as design patterns, architectural tactics, and object-oriented principles like inheritance and encapsulation \cite{saad2025senai}. Most LLM development to date has focused on models that generate functionally and syntactically correct code \cite{haque2025llms}. This creates a gap between the capabilities of these models and the broader expectations of software engineering practice. For example, Wang et al.~\cite{wang2024oop} showed that current LLMs struggle with Object-Oriented Programming (OOP) concepts. As a result, these models may produce poorly structured software designs or designs that do not consider the reuse of existing source code. Such designs can lead to systems that are difficult to maintain and more prone to accumulating redundancies and technical debt. 

Without additional measures like Retrieval-Augmented Generation (RAG) or GraphRAG \cite{han2025graphrag}, GenAI models also struggle to understand company-internal context and generate suitable output due to their lack of relevant training data. There are two potential solutions to this: fine-tuning the LLM to internalise the domain-specific knowledge or provide the correct context to garner a specific response. Both of which suffer from similar problems surrounding knowledge management. Fine-tuning requires knowledge provided through datasets that is accurate, complete, and well-structured to be successful. However, the necessary data is often not in this condition and requires much effort to pre-process and obtain. Likewise, providing high-quality context requires knowledge that meets this criterion along with additional requirements such as availability and high-velocity retrieval. To address these challenges, methods such as In-Context Learning (ICL), RAG, and knowledge graphs as context sources such as those created in GraphRAG, have been developed.  

However, these approaches do not address the critical challenges related to the distribution of logically connected data and data sources, which is often the case in the SDLC (e.g., requirements, user stories, specifications, architecture and design documents, source code, bug reports, and end-user documentation). They also do not address the inconsistent nature of the data required to build knowledge bases. Furthermore, they do not resolve the fact that a large portion of this knowledge may be undocumented. With the proliferation of the Model Context Protocol (MCP) \cite{hou2025mcp}, the integration of external systems is becoming increasingly simpler. However, even if MCP were to be used to integrate with a project management software for instance, the use of it during software development activities may be inconsistent. The effectiveness of the MCP integration for knowledge management is only as good as the data within the underlying systems.

Another prominent issue is the lack of GenAI support for cross-SDLC engagement. GenAI solutions are often narrowly constrained to isolated use cases, with AI artifacts and outputs commonly remaining ‘single-use’ -- either forgotten or inaccessible during later SDLC phases. However, accessible and queryable storage of these artifacts is required as contextual input to enable seamless application of GenAI across different SDLC phases.

Lastly, while context windows of LLMs have supposedly increased to as much as 100 million tokens for models such as LTM-2-mini \cite{ltm2mini}, the resources required, and efficiency of response is yet to be scientifically evaluated. 

\subsection{Security and Data Privacy Risks} \label{sec:security}

LLMs can introduce security vulnerabilities into the generated artifacts such as source code or domain information \cite{bar2025ai}. This poses risks to the security and integrity of the developed software. Furthermore, LLMs can be exploited to support or carry out cyberattacks. 

Several benchmarks exist to evaluate a model’s risk in producing insecure or vulnerable code as well as its potential to facilitate cyberattacks \cite{bar2025ai}. However, most of these benchmarks present several limitations. They do not provide a comprehensive assessment of both insecure coding practices and the model’s potential to enable cyberattacks, often focusing only on code completion or natural language prompts without full end-to-end attack generation \cite{bhatt2023purple}.

Most benchmarks rely on static evaluation methods, such as rule-based checks or LLM judgements, which are less precise than dynamic testing and prone to errors \cite{thakur2025judging}. Additionally, there is a trade-off between dataset quality and scalability: manually created datasets offer high quality but limited scalability, while automated datasets scale better but tend to be lower quality and less relevant to real-world security issues.  

Language also introduces a trade-off, as benchmarks prompting models in natural language may not fully capture the model’s actual ability to generate harmful code, sometimes in a specific language, limiting the realism and depth of the evaluation \cite{jain2024livecodebench}.  

Adversarial attacks targeting code generation models represent another critical aspect that must be considered \cite{yang2022natural}. Various types of LLMs used for code generation are known to be susceptible to such attacks, which can compromise system integrity and security \cite{yang2024seccodeplt}. As a result, incorporating adversarial training into the development of these models is essential to enhance their resilience and protect the security and privacy of the systems in which they are deployed \cite{awal2025large}.

\subsection{Biases in LLM Training Data} \label{sec:biases}

LLMs are trained on vast datasets, which may reflect historical biases or flawed software engineering practices present in the software development community \cite{atemkeng2024ethics}.  

These biases may manifest across sensitive attributes such as age, gender, and education. If the training data contains information that makes gender or race-based assumptions in user data processing, the LLM may learn and perpetuate these biases in its generated outputs, such as code. This can lead to inequitable software solutions and unfair user experiences. Companies using biased datasets and LLMs in software engineering risk generating discriminatory software that can lead to legal and reputational damage. Existing bias testing methods, designed mainly for natural language, struggle to effectively identify biases within the logical structure of code \cite{huang2025bias}.  

Alongside ethical considerations, bias in training data can materialise in several other ways. Bias towards certain technical design choices, such as libraries, vendors, and architectural styles can be perpetuated according to the model supplier, e.g., certain models being trained on significantly more solutions within a particular cloud environment and thus favoring those services when making architectural decisions, rather than considering what is best for the requirements. There is ongoing research that has identified distinct preferences LLMs have for programming languages and libraries \cite{twist2025study}. 

\subsection{Environmental Impact and Sustainability} \label{sec:sustainability}

Despite its promise, GenAI introduces new environmental and structural risks across the SDLC.  These remain under-researched and are rarely addressed in practice. Initial analyses show that GenAI-assisted coding and testing workflows carry significant energy and emissions costs -- particularly when scaled across teams, automated through agents, or used recursively for prompt tuning and test case generation. Luccioni et al.~\cite{luccioni2024environmental} report that deployment of the BLOOM large language model (176 billion parameters) over just 18 days consumed an average of 40 kWh of electricity per day and generated approximately 19 kg of CO\textsubscript{2} emissions per day, depending on the cloud provider's energy mix. This illustrates the substantial operational footprint of GenAI models even after training, reinforcing the need to account for emissions across the full deployment lifecycle. The first challenge in this domain is to provide reliable end-to-end models of carbon footprint (including the embedded component) of the new processes and workflows enabled by GenAI. We currently see early results and research projects emerging in this area \cite{faiz2024llmcarbon,greensoftware}.

The second challenge is to find patterns in AI-assisted processes and workflows that mitigate the environmental sustainability risks. Our early results show that architectural patterns and problem decomposition may be key in finding such sustainable patterns \cite{cheung2025comparative}. Finally, coming up with action plans that facilitate the migration of legacy and AI-assisted processes and code to more environmentally friendly patterns is a contemporary challenge. In all these challenges, GenAI can serve as a key component, once its environmentally sustainable use is well-researched and prioritised.

\subsection{Unexplored Long-Term Impacts of GenAI in the SDLC} \label{sec:long-term}

Evaluating the output of LLMs in software engineering remains a significant challenge. As indicated in Section \ref {sec:context}, existing evaluation methods, especially for code generation, often rely on benchmarks for coding tasks that do not fully reflect real-world software engineering challenges. These include large, industrial-grade codebases with complex, evolving technology stacks and company-specific programming frameworks and libraries. Additionally, many benchmarks do not assess whether the generated code adheres to good software design practices. Consequently, models may achieve high evaluation scores even when the generated code is poorly structured or fails to follow essential software engineering principles. 

Beyond the evaluation of LLM-generated output, long-term assessments of GenAI's impact on the SDLC are still scarce. While many studies have explored the short-term effects of GenAI-based coding assistants \cite{cihan2024automated}, there is limited research on their influence on long-term industry metrics, such as those defined by the DORA framework. Initial findings from the 2024 DORA report \cite{dora2024} suggest that GenAI improves capabilities typically associated with better software delivery performance, but also highlight declines in areas such as software stability and throughput. 

Longitudinal research into 2025, however, reveals a complex and evolving landscape where teams and tools have adapted. The 2025 DORA report \cite{dora2025} suggests that AI acts as an amplifier, magnifying existing organisational strengths and weaknesses. For example, while AI adoption improves software delivery throughput and enhances individual effectiveness, it also increases software delivery instability, indicating that underlying systems have not yet evolved to safely manage AI-accelerated development. 

The increased adoption of AI-authored code can lead to a significant surge in duplicated code blocks, often without developers' explicit awareness. This creates substantial challenges for the long-term maintainability of source code. At the same time, AI's influence appears to be correlated with a decline in code refactoring activities, which are essential for consolidating existing work into reusable modules and minimising the number of systems requiring maintenance \cite{gitclear2025ai}.  

More research is needed to understand the long-term impact of GenAI on software engineering practices.

\subsection{Evolving Software Engineering Roles and Processes} \label{sec:roles}

With the expanding influence of GenAI across the SDLC and its acceleration, the role of software engineers will be evolving. Developers must face these challenges and acquire and strengthen different competencies than what they have been trained for, including prompt and context engineering, AI oversight, debugging AI-generated code, and managing AI models. These skills are needed to effectively guide AI tools and ensure the integrity of the software they help produce. 

In addition to up-skilling challenges, over-reliance on AI tools, particularly when developers blindly adopt suggestions, can erode critical thinking and creativity \cite{campos2024things,mastropaolo2024rise}. Junior developers may miss out on foundational learning, creating skill gaps within teams. This can lead to situations where only senior engineers understand the system's architecture and design decisions \cite{stackoverflow2024survey}. Moreover, AI-generated code is not always optimal, contextually appropriate or secure (as described in Section \ref{sec:security}). It often lacks documentation and may follow unconventional patterns, making it harder for engineers to interpret, maintain, and modify the code. These obstacles can increase technical debt and long-term development costs. 

Likely in the short term, we will see software development be performed in line with best practice before the arrival of artificial intelligence. However, the assistance of AI will lead to a shift in the focus of activities. Developers will progressively move away from active coding and instead focus on breaking down complex coding tasks for AI tools while ensuring the quality of the generated output. In this context, AI coding assistants already help reduce the time required to create minimum viable products. However, this also introduces challenges in managing unprecedentedly large codebases developed within very short time spans. Some of the authors observed this first-hand in their company context, where the amount of code increased by a factor of two to three within the time span of a year. A new process of engineering will be needed (as outlined in Section \ref{sec:selfsystems}) to deal with this unprecedented pace.  

For the future, we foresee significant changes in traditional activities currently considered fundamental to the development process. As AI tooling advances towards achieving a critical level of reliability and robustness, many of these activities may become obsolete. Specific programming tasks can be fully automated, testing and quality assessment can largely be performed autonomously, and project management and delivery can be coordinated across teams and organisations\cite{beheshti2024natural,hassan2024ainative}. At this point, organisations will be challenged to reconsider their development and business processes in novel ways to take maximum advantage. For example, feature updates or bugfixes in well-understood areas can be delivered autonomously rather than go through human-managed processes to enhance speed of delivery, something currently unthinkable.

\section{The Future of GenAI-Driven Software Engineering: Our Vision} \label{sec:vision}

Over the next five years, we anticipate significant advancements of AI in software engineering that will address the limitations highlighted in Section \ref{sec:challenges}. Drawing on our experience and findings, this section outlines our vision for how these technological evolutions will transform the software development process.

\subsection{Towards Self-* Systems} \label{sec:selfsystems}

Most software engineering artifacts along the SDLC, code in particular, will be generated by AI at an accelerated pace \cite{qiu2025todays,jin2025advancing, hassan2024ainative}. Software engineering tasks such as requirements gathering, architectural design, and testing are becoming increasingly automated, especially through using agentic AI solutions \cite{jin2024mare,jin2024llms}. This will increase the capacity of software engineering teams and enable them to focus on innovative problem-solving. As AI takes on routine tasks such as coding and testing, software engineers will play a critical role in orchestrating AI efforts by breaking down complex problems into manageable components, just as mentioned in Section \ref{sec:roles}. In this context, maintaining sustainable minimalism, particularly in code, becomes essential to ensure long-term readability, maintainability, and architectural integrity \cite{fowler2024panel}. 

To cope with the automated and accelerated development process, engineering teams must establish the processes and workflows to validate, verify, and optimise AI-generated artifacts to ensure software quality -- particularly in industrial-grade systems where reliability, maintainability, and performance are paramount \cite{terragni2025future}. Automation and rigour in these workflows are essential to mitigate the risk of fatigue and burnout \cite{alami2025human} due to continuous human oversight. Personalised feedback will be built into these processes to facilitate decision-making and oversight. This is also particularly important due to the numerous security risks documented in Section~\ref{sec:security}, such as the introduction of security vulnerabilities into code, or the utilisation of LLMs within software providing additional attack vectors.  

The increasing autonomy and reliability of GenAI within the SDLC establishes the foundation for the emergence of genuinely self-* systems. A self-* system is a system that can alter itself at runtime, without human intervention, and encompasses the concepts of self-aware, self-healing and self-evolving systems \cite{salehie2009self,beheshti2024natural,hassan2024ainative,jin2024mare}. These factors will coalesce to form the next generation of self-* systems. They are self-evolving, considering changes in requirements from various human-approved external sources, such as public reviews and bug reports, having contextual information about architectural structures and redeveloping it, testing and deploying themselves to meet these new requirements. All of these will be facilitated by the AI autonomous agents and novel forms of agentic teams. It is through the evolution of these agentic teams that we can address the challenges with limited reasoning capabilities and the proliferation of hallucinations, as mentioned in Section \ref{sec:hallucinations}. Agents working together autonomously for a shared purpose will validate each other’s outputs and pool their individual expertise to tackle harder problems. As an analogy, the evolution of AI in software engineering towards self-* systems can be likened to the progression of driver-assistance systems evolving into fully autonomous driving -- a journey from supportive automation to complete autonomy. 

Part of this self-evolution process will be the enhanced facilitation of research spikes, experimentation and A/B testing within a system. A common yet costly practice now could be much more efficient in the future, with agents automatically identifying, designing, implementing, testing and concluding alternative approaches. 

Once a system’s requirements are stable, these self-* systems can also become self-healing. Traditionally, this has been defined as being able to adapt to different environmental conditions and continue to behave as previously specified, typically associated with automatic scaling of components to deal with varying loads, e.g., in cloud deployments. However, future AI agents will allow for automatic detection of faults and runtime errors, modifying the system at the source code and at the architectural level, to patch said faults, all while meeting the same underlying requirements. 

Even in the context of self-* systems, AI will not fully replace the need for human creativity, judgment, and domain expertise. Maintaining a balance between leveraging AI and cultivating human talent will be essential for sustaining innovation, quality, and resilience in the software industry \cite{gartner2024ai}.

\subsection{Intuitive, Inclusive Multi-Modal Human-AI Collaboration} \label{sec:human-ai}

As mentioned in Section \ref{sec:selfsystems}, the future will witness the progression beyond this current state of the art with multiple agentic teams, each with their own roles, tools, and objectives, working across these SDLC phases to autonomously develop and deliver software systems end-to-end. Their interactions and relations will be loosely defined, driven by the agents and their internal decision-making processes rather than by pipelines, while being self-governed by their internal ethics and human-defined regulations. Along with the progressive shift towards autonomous development, we will begin to see a shift in interaction patterns, as teams interact with AI through a diverse range of modalities beyond the traditional keyboard, such as voice, touch, and gesture-based interactions. This will allow for a dynamic workflow that best fits the constraints of engineering teams while remaining within the bounds of safety and law. 

In essence, humans will interact with the development process through natural language that centres around the description of a system through requirements and features, sitting much closer to the end users of a system and how they perceive it. This will likely be utilised alongside the more "traditionally no-code" approach of visual coding, defining a system through how the user interface should look and function. 

As this approach will be akin to natural language, it will lend itself better to voice instructions. This interaction pattern will suit certain developers better than others, but will certainly improve the experience for those with specific disabilities and impairments, such as sight impairment or dyslexia. 

Touch gestures and electronic pens can be used to express intent and modifications – for example on code. The AI would provide real-time feedback on its interpretation, enabling iterative refinement without extensive textual prompting. This approach would leverage visual thinking for a more direct and less abstract coding experience. During group-based brainstorming and modelling sessions, a virtual whiteboard can be augmented with a GenAI-based assistant to explore group ideation with AI in the loop \cite{he2025llmbased}. 

The more proactive agents previously discussed would contribute in much the same way as other humans, listening to conversations and providing thoughts and ideas. Subsequently, AI-based engineering systems could consume the multimodal outputs of this interaction, such as speech, architecture diagrams, or wireframes, to gain further insight into the design process and produce better code outputs. These group ideation scenarios often unintentionally isolate those who communicate non-verbally, such as sign language users. Utilising AI for understanding and audibly involving them will also contribute to better, more efficient collaboration. 

Humans will continue to work alongside these agentic teams, with inter-human collaboration becoming streamlined. Data will become more up-to-date through rapid task execution, more complete and transparent due to the agents' self-documentation of every decision point and the reasoning behind it, and more queryable than ever due to agentic knowledge retrieval. 

This vision will be enabled through the maturation of context and data management within organisations. As mentioned in Section \ref{sec:context}, limited contextual awareness severely hinders an AI system’s ability to produce meaningful outputs. Organisations will address this by enhancing the rigour with which they collect, store, and interpret data, driven by the aspiration to leverage this data with AI, particularly autonomous AI agents. The agents themselves, when working with data, will therefore also identify poor-quality data during operation and either propose improvements or independently implement enhancements. This trend towards a robust and accurate representation of a 'single source of truth' will facilitate the rich, ad-hoc collaboration of an agentic ecosystem in software engineering. 

Moreover, most agents will become more proactive, acting on changes in their software engineering environment without being manually invoked. This proactivity will be made possible by possessing an accurate, agent-centric representation of the situation through this single source of truth. 

\subsection{Improved Trustworthiness: Reasoning, Robustness and Transparency} \label{sec:trust}

We envision incorporating several key enablers to increase the trustworthiness of AI in software engineering, including stronger reasoning capabilities, greater robustness, and improved explainability and transparency. 

There is ongoing research on expanding the reasoning capabilities of GenAI with other reasoning components, for example based around inference engines, world models and ontologies. These mechanisms will allow modelling of foundational software engineering principles like architectural design patterns and tactics \cite{saad2025senai}. We anticipate that the integration of such knowledge during the training phase will enable LLMs to not only generate code that adheres to software engineering best practices, but also assist with higher-level tasks such as architectural analysis, performance optimisation and security aspects. These techniques, along with those mentioned in Section \ref{sec:selfsystems}, will contribute to the mitigation of challenges discussed in Section \ref{sec:hallucinations}. 

Additionally, the maturation of a model’s chain-of-thought is integral to improving robustness. Currently, chain-of-thought monitoring aims to ensure that if a model starts to pursue an unrelated topic that will negatively impact the correctness of the answer, it can be terminated and restarted, even re-prompted. However, current approaches rely on having a natural language representation of said chain-of-thought. New approaches are currently being developed to improve the efficiency of models utilising chain-of-thought by using non-human-readable languages, often dubbed “Neuralese” \cite{hao2024training}. Ensuring that chain-of-thought monitoring methods can work alongside these new approaches is essential.

A key factor for trustworthiness is more increased insight into the training data and process that led to the creation and behaviour of the LLMs used. For example, Lindsey et al.~\cite{lindsey2025biology} apply attribution graphs to trace how an LLM arrives at particular outputs, thereby offering one avenue towards understanding model-internals and improving transparency. Additionally, the architecture around GenAI should provide truthful and understandable information about how a GenAI model arrived at the provided answers. Such mechanisms should be incorporated both in the architecture of the AI infrastructure and into the SDLC process.  

As mentioned in Section \ref{sec:biases}, the various types of bias held by LLMs primarily derive from the contents of the training data. The only way in which these biases can be reduced is by having a thorough understanding of the training data. We estimate that once the hype cycle surrounding GenAI progresses and subsequently wanes, consumers and lawmakers will demand greater transparency from model providers, forcing them to reveal the underlying training data \cite{euaiact2025transparency}. Once this occurs, establishing evaluation techniques and benchmarks that assess bias at the training data level, opposed to the current method of a standard question-answer benchmark, will give further insight into the origins of the bias, enabling more effective bias mitigation.

\subsection{Embedding Security and Data Privacy into an Autonomous SDLC} \label{sec:emb-security}

For the future, we anticipate that security and data privacy will be significantly incorporated into AI models and their infrastructure. Advancements, e.g., initiated by frameworks like TRiSM \cite{litan2024tackling} and ENISA \cite{polemi2023multilayer} are heading towards structured defenses organised around trust, risk, and security. These frameworks emphasise continuous monitoring, algorithmic scoring, automated categorisation as well as human oversight in line with the EU AI Act \cite{euaiact2024}. 

One of the primary security challenges identified in Section \ref{sec:security} was the proliferation of security vulnerabilities in AI-generated code from the AI’s training data. The agentic teams discussed in Section \ref{sec:selfsystems} can also serve to be the resolution to this challenge. It cannot be guaranteed that all security vulnerabilities are found and removed from the vast amounts of training data. The agents responsible for generating code will also be able to autonomously analyse the code they produce, employ state-of-the-art static and dynamic code analysis tools and methods to detect vulnerabilities and patch them, as well as execute various testing frameworks to validate the generated code. 

Benchmarks used to assess the security of AI-generated code will also become more robust, employing dynamic techniques such as sandboxed executions of code to provide a more in-depth, true-to-life determination of the LLM’s code generation abilities. Future benchmarks will evaluate both the security of the generated code and the AI's ability to resist being used in cyberattacks, such as creating malware or acting as an attack tool for autonomous actions. 

Advances in AI-augmented security testing and lifecycle integration amplify these autonomous protection capabilities. Static, dynamic, and hybrid application security testing are increasingly coupled with emerging LLM-based reasoning engines such as Codex Security Scanner \cite{voskanyan2025codexsentinel} and SecCoder GPT \cite{zhang2024seccoder}, which apply contextual understanding to detect vulnerabilities beyond pattern-based tools.

Privacy-preserving mechanisms, including federated learning or homomorphic encryption, further strengthen confidentiality, while differential privacy safeguards individual data contributions. Complementary threat modelling frameworks like STRIDE or PASTA \cite{naik2024comparative} systematically map risks across components and phases, guiding mitigation of spoofing or adversarial manipulation.  
However, embedding traditional secure coding standards from OWASP or the ML Top 10 \cite{owasp2023mlsec} will remain crucial to ensure early and consistent enforcement. 

\subsection{Embedding Sustainability into GenAI-Driven Software Engineering} \label{sec:emb-sustainbability}

To address the challenges raised in Section \ref{sec:sustainability}, we propose a vision in which sustainability is embedded as a measurable, actionable objective throughout the GenAI-based software development process. Rather than treating environmental impact as a downstream effect, sustainability will be built directly into the design, generation, and evaluation stages of AI-assisted engineering. 

We envision an approach that is grounded in two core principles. First, sustainability must be treated as a requirement across the SDLC. Second, GenAI systems must be able to respond to sustainability-related constraints -- generating code and decisions that are not just accurate or performant, but efficient, reusable, and low-impact by design. 

Our vision will be operationalised through various changes to process and developer opinion. Baseline emissions profiling of GenAI-assisted workflows should become commonplace, as without an accurate assessment of the current environmental impact, there is no way of improving it effectively. We already see early examples of this part of our vision being implemented in practice \cite{faiz2024llmcarbon,cheung2025comparative,codecarbon2025}.  These measurements should be taken and monitored inherently as part of the SDLC, just like other metrics associated with software quality. At the architecture and source code level, embedding sustainable design practices such as duplication and complexity minimisation will ensure that systems remain as sustainable as possible as they mature.  

Alongside, the way in which we interact with LLMs should also be optimised for environmental impact. We envision that certain workflows and processes are likely to produce higher quality software with reduced carbon footprint \cite{cheung2025comparative}. Modifying operationalisation of GenAI to reduce the energy required to produce a certain output will become the standard, in addition to hyperscalers and model providers optimising the underlying platforms in the same way. 

Evidence from sustainable computing research supports this approach: energy consumption in code varies depending on programming language, algorithmic structure, and how system resources are accessed. Danushi et al.~\cite{codecarbon2025} report that choosing suitable algorithmic strategies and data structures can help reduce the number of operations, improve memory access patterns, and ultimately lower energy consumption. If GenAI systems are guided towards such optimisations -- through prompting, constraints, or agentic self-assessment -- they can reduce environmental impact without compromising productivity. 

This vision also aligns with wider movements in responsible AI and environmental governance. Standards such as ISO/IEC TR 30165 \cite{isoiec30165} and initiatives like the UN’s Green Digital Action Pillar \cite{greendigital2023} call for systems that are not only functional, but measurable, transparent, and accountable.

\subsection{Evolution of Software Engineering Roles and Skills} \label{sec:evol-roles}

As outlined in the previous sections, GenAI increasingly becomes a natural part of software engineering along the entire SDLC, for which we envision the potential impacts of this on software engineering activities and the required skills. The changes to the skills expectations can already be seen as a result of the roll-out of coding assistants such as ChatGPT \cite{chatgpt}, GitHub Copilot \cite{copilot} and Cursor~\cite{cursor}. In this section, we will also examine the longer-term context of agent-based development systems and end-to-end AI support for software engineering. 

As discussed in Section \ref{sec:selfsystems}, the direct impact of coding assistants can already be observed as they streamline a significant amount of development effort while at the same time increasing effort in areas such as code reviews~\cite{google2025q3}. While at this point this does not fundamentally change any roles, it marks a shift in where the majority of the effort is spent when developing software. A major part of the work becomes prompting and reviewing as discussed in Section \ref{sec:roles}. This requires a good understanding of the capabilities of the various models and the problem space to decide both whether to rely on LLM-generated code for addressing the problem at hand as well as making optimal use of it.  

This work will therefore likely require a higher technical competence level grounded in deep domain and technology understanding, which is typically associated with more senior (as opposed to junior) developers.  Additionally, using GenAI will require a high degree of metacognitive monitoring and control -- the psychological ability to monitor and guide one's own thought processes -- to adopt to challenges in prompting, output evaluation, and workflow automation strategies \cite{tankelevitch2024metacognitive}.  

A longer-term and more pervasive impact will likely be the results of agentic assistants supporting the SDLC (Section \ref{sec:human-ai}). These agentic capabilities will have increased autonomy (Section \ref{sec:selfsystems}) and will focus on tasks beyond coding, incorporating additional steps such as requirements engineering \cite{jin2024mare}, planning and task management, and quality assurance. This increased autonomy and overall increase in accuracy compared to traditional GenAI tooling \cite{kwa2025measuring} mean that we can anticipate coherent work results across many hours, if not days, of effort in the future.  

This likely will result in shifts of developer effort into other disciplines (requirements engineering, architecting) towards a more review-oriented execution, requiring more critical analysis and breadth of knowledge to assess the proposed solutions.  

While for critical processes and specialised topics (like high-performance computing, embedded systems and mission-critical control systems) a human in the loop is still required \cite{saifi2025let}, it is within the realm of possibility that more well-understood tasks can be fully automated with agentic AI, which will lead to a reduced need for the skills human require to perform them.   

A prominent challenge facing organisations in the future is that the remaining tasks left in future development are primarily done by senior people. This shift threatens the competence acquisition pipeline and alternatives need to be devised. 

The logical conclusion for AI support is to have integrated support across the development lifecycle, an approach of which early signs of success have been seen \cite{wikipedia2025vibecoding}. In the future, agent-based approaches scale to more complex tasks with high quality across activities, with these systems potentially replacing complete workflows across teams. In such situations, humans will shift from an engineering towards an orchestration perspective, i.e. they will be users with a sufficient understanding of technology, able to supervise, manage, and correct development processes. Effectively, this means people need to have competences that are currently typical of product managers and technical leads. At the same time, new roles will emerge that support and enable these AI capabilities to function across an organisation, such as knowledge capture and management, new AI capability creation and AI governance. And of course, this also comes with broader skill expectations when building software systems that incorporate GenAI. Technical skills, such as retrieval-augmented generation, embeddings, vectorisation, specific AI frameworks like LangChain \cite{langchain2024} and AWS Strands \cite{amazonq}, and AI-specific communication protocols like MCP, A2A will be increasingly essential. Enhanced data-thinking skills will also be required, including critically evaluating data sources, identifying hidden biases, formulating precise analytical questions, and translating insights into robust and ethical software solutions \cite{mike2022computational}. A core challenge for this scenario will lie in skill acquisition and retainment especially for companies that rely on outsourcing: while separate skills are required to address the novel challenges, it is difficult to acquire and retain the necessary depth of skills in an organisation to adequately judge developed solutions. This comes on top of the issue of a shift towards more senior competences.

Addressing these shifts demands deliberate strategies for continuous learning and organisational adaptation, ensuring that human expertise, motivation, and engagement remain vital complements to increasingly autonomous AI systems.

\section{Outlook on Implementing the Vision in the GENIUS Project} \label{sec:outlook}

Building on the outlined current challenges and envisioned future directions, we now turn to the practical realisation of this vision. The ITEA4 GENIUS project \cite{itea-genius} brings together leading European industrial and academic partners to develop and evaluate key technological and methodological foundations, translating this broader vision into tangible, validated GenAI-based solutions across the SDLC. 

\subsection{Overview of the GENIUS Project}

The ITEA4 GENIUS project ("Generative AI for the Software Development Life Cycle") is a European research and innovation initiative that brings together more than 30 industrial and academic partners from 8 countries: Germany, Austria, the United Kingdom, Türkiye, Portugal, Finland, Belgium and Canada. The consortium comprises 10 large industrial enterprises, 12 small and medium-sized companies, and 9 universities and research institutes, representing a wide spectrum of domains such as industrial manufacturing, telecommunications, automotive, healthcare and IT services. The project’s overarching objective is to advance scalable, trustworthy, and effective GenAI-enabled software engineering across all phases of the SDLC. 

GENIUS aims to leverage the potential of GenAI to enhance and automate complex software engineering tasks,  ranging from requirements analysis and code generation to documentation and quality assurance. It explicitly addresses the challenges of applying these technologies in industrial  contexts, including their integration into complex software ecosystems, assurance of model quality, and compliance with domain-specific standards. By systematically combining applied research, tool development, and industrial evaluation, GENIUS aims to establish the foundations for GenAI solutions that are not only technically sophisticated but also production-ready and aligned with real-world requirements. 

GENIUS is characterised by its scale, industrial grounding, and holistic perspective on the software lifecycle. The project conducts large-scale studies and experiments with real developers, codebases, and organisational processes across multiple European industries, creating a rich, empirically founded evidence base. Its lifecycle-centric approach covers all phases of the SDLC, revealing integration and traceability challenges that are often overlooked in isolated evaluations. Moreover, GENIUS adopts an ecosystem-oriented view that extends beyond model performance to include knowledge management, development workflows, and human oversight, laying the groundwork for trustworthy and effective GenAI adoption in industrial environments. 

In line with the strategic objectives of European technological sovereignty, the GENIUS consortium places a strong emphasis on ensuring that the development and deployment of GenAI solutions remain aligned with European values, regulatory frameworks, and innovation capabilities. This includes actively prioritising and tracking the adoption of European-based GenAI infrastructure, such as cloud service providers and open-source LLMs like Mistral and other emerging initiatives. GENIUS also seeks to align closely with evolving policy frameworks, in particular the EU AI Act, by emphasising transparency in training data, traceability, and ethical AI principles. Through this, the project contributes not only to technical innovation but also to shaping a responsible and self-determined AI landscape for European software engineering. 

\subsection{Use Cases as Drivers for Industrial Relevance and Evaluation}

The GENIUS project is structured around 14 industrial use cases that serve as the foundation for iterative tool development, testing, and evaluation. These use cases, defined and maintained by key industrial partners, are instrumental in ensuring that project outcomes are directly relevant to real-world development practices. They not only define technical requirements and deployment environments but also offer a testbed for the integration and validation of project innovations. 

Among the leading industrial contributors, Siemens AG contributes to the GENIUS project with a strong emphasis on enhancing developer experience through GenAI-driven services embedded in the Siemens Developer Portal \cite{siemensdev}. This use case explores how intelligent assistants and agents can support software teams by offering context-aware guidance, automated documentation, and code-related services tailored to specific project phases and domains. The goal is to create a seamless, AI-augmented development environment that empowers software engineers with timely insights, reduces cognitive load, and fosters higher productivity and quality throughout the SDLC. These innovations are designed to integrate smoothly into Siemens’ existing developer ecosystem and address key industry requirements for scalability, reliability, and traceability. 

British Telecom (BT) contributes use cases that centre around three major themes: architectural design, value quantification of GenAI and knowledge management. The first is concerned with embedding GenAI into the architectural design process, hoping to explore themes of software quality optimisation and automated architectural refactoring. The second will see the development of a common framework for the quantification and representation of GenAI impact for SDLC use cases when deployed within organisations. The final theme concerns the advancement of knowledge management and utilisation methods that power multi-agent systems. BT has already demonstrated measurable improvements in productivity and software maintainability through GenAI-supported workflows and will further expand these tools in GENIUS. 

Akkodis, a global engineering services provider, contributes multiple use cases that emphasise automated ticket resolution and the automatic generation of test specifications and test programmes. Their approach leverages GenAI to scan and analyse historical data, extract knowledge and create search agents, and automate the derivation of test code from high-level requirements. These capabilities directly address key pain points in quality assurance and operational efficiency. 

The remaining use cases, ranging from embedded systems development to large-scale platform engineering, are distributed among other industrial partners across sectors such as automotive, industrial automation, telecommunication, healthcare, and IT services. They collectively ensure broad coverage of SDLC phases and industry contexts. Together, the use cases serve as the practical backbone for verifying the relevance, scalability, and generalisability of GENIUS innovations. 

\subsection{Technological Innovations Addressing Core Challenges}

The technological developments in GENIUS follow an integrated approach that connects all major phases of the traditional SDLC model. Advanced GenAI solutions for requirements analysis, system design and code development, and quality assurance are developed in close interaction, with the resulting methods, tools, and models continuously integrated into a shared infrastructure that ensures evaluation, interoperability, and reuse. Artifacts created in earlier phases serve as input for later validation and optimisation, and testing results and feedback are iteratively transferred back to refine upstream processes. This bidirectional flow of knowledge and artifacts ensures consistency, traceability, and continuous improvement across the lifecycle, consolidating all developments through a common technological backbone that advances the overall vision of reliable, context-aware, and integrated GenAI-driven software engineering. 

\vspace{4pt}
\textit{Requirements Analysis and Documents Processing:}
This innovation area advances requirements engineering through AI-driven document intelligence, transforming heterogeneous industrial artifacts such as specifications, technical reports, standards, and project documentation into structured, validated, and reusable knowledge. Building on advanced LLMs enhanced with retrieval-augmented generation, semantic search, and domain ontologies, GENIUS develops pipelines that automatically identify and extract requirement-related information, generate stakeholder-specific summaries, and harmonise diverse document sources into coherent, traceable representations. 

The developed assistants will not only extract and refine existing requirements but also generate and adapt new ones from text, chat, audio, or code, thereby enriching incomplete specifications. Human-in-the-loop validation ensures interpretability, security, and fairness, while interactive features allow engineers to clarify, correct, or extend AI-proposed content dynamically. Further innovations address requirement analysis and verification, where models evaluate completeness, consistency, and clarity by detecting contradictions, ambiguities, and missing information, supported by readability scoring, terminology alignment, and visual feedback for collaborative review. 

Together, these technologies establish self-improving and explainable requirement pipelines that embed domain expertise, enable cross-phase traceability, and form a reliable foundation for downstream design and testing. They directly address the challenges of hallucination and limited reasoning (\ref{sec:hallucinations}), insufficient contextual grounding (\ref{sec:context}), and bias in generated content (\ref{sec:biases}), contributing to the vision of trustworthy, context-aware, and collaborative GenAI-driven software engineering (\ref{sec:selfsystems}-\ref{sec:trust}). 

\vspace{4pt}
\textit{System Design and Code Development:}
Building on the structured and validated requirements generated in the previous phase, this innovation area extends GenAI support towards architecture design, code generation, and software comprehension. GENIUS develops intelligent assistants that integrate architectural reasoning, reusable design patterns, and domain knowledge to ensure that system structure and implementation remain consistent, explainable, and verifiable across the lifecycle. The approach directly tackles the challenges of limited reasoning and structured output (\ref{sec:hallucinations}) and insufficient contextual awareness (\ref{sec:context}) by grounding generation in explicit architectural context and applying grammar- and type-constrained decoding to enhance correctness and coherence. 

Central to this work are architecture-aware generation pipelines, where LLMs interact with analytical components to derive design views, recommend suitable architectures, and generate high-quality code templates from validated requirements and existing artifacts. Generated code is continuously analysed for quality, performance, and security (\ref{sec:security}), embedding secure-by-construction principles and automated detection of vulnerabilities and dependencies (\ref{sec:emb-security}). In parallel, the systems support reverse engineering and model co-evolution, extracting architectural insights from existing implementations to ensure alignment between design intent and realised code, thereby addressing long-term maintainability and technical debt (\ref{sec:long-term}). 

To enhance robustness and reliability, GENIUS integrates multi-agent collaboration frameworks, in which specialised agents for design, coding, and evaluation cross-validate each other’s outputs. By uniting creative generation with systematic validation, this work advances the vision of robust, autonomous, and trustworthy GenAI-driven software engineering (\ref{sec:selfsystems}-\ref{sec:trust}) and forms an essential bridge between requirements understanding and quality assurance in the overall SDLC. 

\vspace{4pt}
\textit{Quality Assurance and Maintenance:}
This innovation area strengthens the reliability and resilience of GenAI-assisted software development through automated testing, defect analysis, and continuous validation. GENIUS develops intelligent assistants and pipelines that transform requirements, designs, and code artifacts from earlier stages into executable test cases, quality metrics, and optimisation strategies. By embedding validation directly into the development loop, the project tackles the challenges of hallucination and limited reasoning (\ref{sec:hallucinations}) and security and privacy (\ref{sec:security}), ensuring that AI-generated outputs are consistently verified against specifications and operational data. 

The technological core focuses on AI-enhanced test generation and prioritisation, where models derive and adapt test cases from structured requirements and code to maximise coverage and detect inconsistencies early. Defect and log analysis leverage natural language processing to link issues back to their root causes and recommend corrective actions, while sustainability-oriented optimisation (\ref{sec:sustainability}) reduces resource consumption during test execution. Continuous feedback from testing and maintenance phases is propagated upstream to refine code generation and requirements processing, fostering self-improving quality loops across the SDLC. 

Through these innovations, GENIUS establishes autonomous and explainable quality assurance workflows that close the gap between development and validation. They directly contribute to the vision of sustainable, self-correcting, and trustworthy GenAI-driven software engineering (\ref{sec:selfsystems}-\ref{sec:trust}, \ref{sec:emb-sustainbability}) and ensure that GenAI-based systems remain dependable and adaptable over their full lifecycle. 

\vspace{4pt}
\textit{Cross-Cutting Technological Innovations:}
Complementing the domain-specific innovations in requirements, design, and quality assurance, this area consolidates the technological foundations that ensure interoperability, security, and scalability across all GENIUS developments. The work integrates methods for data preprocessing, model adaptation, and benchmarking into a shared infrastructure, while fostering agile collaboration between research and industry. These cross-cutting activities provide the connective layer of the project, enabling modular, reusable, and privacy-compliant GenAI solutions applicable throughout the SDLC. 

A central outcome is context-aware artifact generation, where requirements, architectural descriptions, and code bases are transformed into executable artifacts such as test cases, models, and quality metrics. By leveraging neural code generation, semantic parsing, and traceability mechanisms, these techniques improve automation and reduce human workload without compromising correctness, compliance, or transparency. This directly addresses challenges of limited reasoning and correctness (\ref{sec:hallucinations}), security and data protection (\ref{sec:security}), and sustainability (\ref{sec:sustainability}), contributing to the vision of trustworthy, self-improving engineering ecosystems (\ref{sec:selfsystems}-\ref{sec:trust}). 

Another major line of innovation concerns semantic search and knowledge management, enabling efficient access, reuse, and integration of distributed engineering knowledge. GENIUS develops graph-based retrieval-augmented generation, fine-tuned LLM pipelines, and domain-specific embedding models to connect requirements, designs, code, tickets, and documentation through semantic links. These solutions form the foundation for intelligent agents with autonomy in SDLC tasks, enhance organisational learning and mitigate knowledge loss, supporting the long-term vision of context-intelligent and continuously evolving development processes (\ref{sec:selfsystems}-\ref{sec:trust}). 

Finally, AI-supported collaboration and cross-phase recommendations extend the use of GenAI towards integrated decision support across SDLC stages. Assistants learn from structured and unstructured data to provide actionable guidance, such as design pattern selection, test prioritisation, or configuration tuning, bridging the gap between isolated development activities. In doing so, they reinforce human-AI collaboration (\ref{sec:human-ai}) and strengthen feedback loops that drive quality, adaptability, and efficiency throughout the lifecycle. Collectively, these foundations not only strengthen present-day engineering workflows but also support the broader societal and organisational transition towards AI-integrated development cultures envisioned in (\ref{sec:evol-roles}). 

Across all these innovations, GENIUS embeds robust safeguards for reliability, privacy, and explainability, including on-premise deployment, fine-tuning under privacy constraints, provenance tracking, and validation of AI outputs against formal specifications. These measures ensure that the project’s breakthroughs are not only technologically ambitious but also industrially viable and aligned with the broader vision of secure, transparent, and sustainable GenAI-driven software engineering. 

\section*{Acknowledgment}

We thank all GENIUS consortium partners for their valuable contributions.
We also thank all reviewers for their constructive and helpful comments, which have contributed to improving the quality of this paper.

During the development of this work, the authors used Google NotebookLM and ChatGPT to enhance the readability and language of the text. After using these services, the authors carefully reviewed and revised the content as needed. The responsibility for the content of this publication lies with the authors.

\bibliographystyle{IEEEtran}
\bibliography{references}

@online{copilot,
  year             = {2025},
  author           = {{GitHub Inc.}},
  url              = {https://github.com/features/copilot},
  title            = {{GitHub Copilot · Your AI pair programmer}}
}

@online{amazonq,
  author    = {{Amazon Web Services}},
  title     = {{AI Assistant -- Amazon Q -- AWS}},
  year      = {2025},
  url       = {https://aws.amazon.com/q/},
  publisher = {Amazon}
}

@online{claudecode,
  year             = {2025},
  author           = {{Anthropic PBC}},
  url              = {https://www.anthropic.com/claude-code},
  title            = {{Claude Code: Deep coding at terminal velocity}}
}

@misc{jiang2024survey,
  title = {A {{Survey}} on {{Large Language Models}} for {{Code Generation}}},
  author = {Jiang, Juyong and Wang, Fan and Shen, Jiasi and Kim, Sungju and Kim, Sunghun},
  year = {2024},
  number = {arXiv:2406.00515}     ,
  eprint = {2406.00515},
  primaryclass = {cs},
  publisher = {arXiv},
  doi = {10.48550/arXiv.2406.00515},
  url = {http://arxiv.org/abs/2406.00515},
}

@misc{hou2024large,
  title = {{Large Language Models for Software Engineering: A Systematic Literature Review}},
  author = {Hou, Xinyi and Zhao, Yanjie and Liu, Yue and Yang, Zhou and Wang, Kailong and others},
  year = {2024},
  number = {arXiv:2308.10620}                       ,
  eprint = {2308.10620},
  primaryclass = {cs},
  publisher = {arXiv},
  doi = {10.48550/arXiv.2308.10620},
  url = {http://arxiv.org/abs/2308.10620},
}

@misc{zhang2023survey,
  title = {{A Survey on Large Language Models for Software Engineering}},
  author = {Zhang, Quanjun and Fang, Chunrong and Xie, Yang and Zhang, Yaxin and Yang, Yun and others},
  year = 2023,
  publisher = {arXiv},
  doi = {10.48550/ARXIV.2312.15223},
  url = {https://arxiv.org/abs/2312.15223},
  urldate = {2025-10-27},
}

@inproceedings{fan2023large,
  title = {{Large Language Models for Software Engineering: Survey and Open Problems}},
  booktitle = {ICSE-FoSE},
  author = {Fan, Angela and Gokkaya, Beliz and Harman, Mark and Lyubarskiy, Mitya and Sengupta, Shubho and Yoo, Shin and Zhang, Jie M.},
  year = 2023,
  pages = {31--53},
  doi = {10.1109/ICSE-FoSE59343.2023.00008},
  url = {https://doi.org/10.1109/ICSE-FoSE59343.2023.00008}     ,
}

@misc{jin2024llms,
  title = {From {{LLMs}} to {{LLM-based Agents}} for {{Software Engineering}}: {{A Survey}} of {{Current}}, {{Challenges}} and {{Future}}},
  author = {Jin, Haolin and Huang, Linghan and Cai, Haipeng and Yan, Jun and Li, Bo and Chen, Huaming},
  year = {2024},
  number = {arXiv:2408.02479}            ,
  eprint = {2408.02479},
  primaryclass = {cs},
  publisher = {arXiv},
  doi = {10.48550/arXiv.2408.02479},
  url = {http://arxiv.org/abs/2408.02479},
}

@misc{nguyenduc2023generative,
  title = {Generative {{Artificial Intelligence}} for {{Software Engineering}} -- {{A Research Agenda}}},
  author = {{Nguyen-Duc}, Anh and others},
  year = 2023,
  number = {arXiv:2310.18648}                                             ,
  eprint = {2310.18648},
  primaryclass = {cs},
  publisher = {arXiv},
  doi = {10.48550/arXiv.2310.18648},
  url = {http://arxiv.org/abs/2310.18648},
}

@online{itea-genius,
  author  = {{ITEA Office}},
  title   = {{ITEA 4 · Project · 23026 GENIUS}},
  year    = {2025},
  url     = {https://itea4.org/project/genius.html},
}

@misc{prather2024hype,
  title = {Beyond the {{Hype}}: {{A Comprehensive Review}} of {{Current Trends}} in {{Generative AI Research}}, {{Teaching Practices}}, and {{Tools}}},
  author = {Prather, James and others},
  year = {2024},
  number = {arXiv:2412.14732} ,
  eprint = {2412.14732},
  primaryclass = {cs},
  publisher = {arXiv},
  doi = {10.48550/arXiv.2412.14732},
  url = {http://arxiv.org/abs/2412.14732},
}

@misc{kalai2025why,
  title = {Why {{Language Models Hallucinate}}},
  author = {Kalai, Adam Tauman and Nachum, Ofir and Vempala, Santosh S. and Zhang, Edwin},
  year = 2025,
  number = {arXiv:2509.04664}                                ,
  eprint = {2509.04664},
  primaryclass = {cs},
  publisher = {arXiv},
  doi = {10.48550/arXiv.2509.04664},
  url = {http://arxiv.org/abs/2509.04664},
}

@misc{gao2024current,
  title = {The {{Current Challenges}} of {{Software Engineering}} in the {{Era}} of {{Large Language Models}}},
  author = {Gao, Cuiyun and Hu, Xing and Gao, Shan and Xia, Xin and Jin, Zhi},
  year = {2024},
  number = {arXiv:2412.14554}  ,
  eprint = {2412.14554},
  primaryclass = {cs},
  publisher = {arXiv},
  doi = {10.48550/arXiv.2412.14554 },
  url = {http://arxiv.org/abs/2412.14554},
}

@article{chen2025deep,
  title = {Deep {{Learning-based Software Engineering}}: {{Progress}}, {{Challenges}}, and {{Opportunities}}},
  author = {Chen, Xiangping and others},
  year = {2025},
  journal = {Sci. China Inf. Sci.},
  volume = {68},
  eprint = {2410.13110},
  pages = {111102},
  issn = {1674-733X, 1869-1919},
  doi = {10.1007/s11432-023-4127-5},
  url = {https://doi.org/10.1007/s11432-023-4127-5}   ,
}

@misc{tie2024llms,
  title = {{{LLMs}} Are {{Imperfect}}, {{Then What}}? {{An Empirical Study}} on {{LLM Failures}} in {{Software Engineering}}},
  author = {Tie, Jiessie and others},
  year = {2024},
  number = {arXiv:2411.09916   },
  eprint = {2411.09916},
  primaryclass = {cs},
  publisher = {arXiv},
  doi = {10.48550/arXiv.2411.09916 },
  url = {http://arxiv.org/abs/2411.09916},
}

@inproceedings{netz2024using,
  title = {Using {{Grammar Masking}} to {{Ensure Syntactic Validity}} in {{LLM-based Modeling Tasks}}},
  booktitle = {MODELS Companion '24},
  author = {Netz, Lukas and Reimer, Jan and Rumpe, Bernhard},
  year = {2024},
  pages = {115--122},
  publisher = {ACM},
  doi = {10.1145/3652620.3687805},
  url = {https://doi.org/10.1145/3652620.3687805}      
}

@misc{park2024grammar,
  title = {Grammar-{{Aligned Decoding}}},
  author = {Park, Kanghee and others},
  year = {2024},
  number = {arXiv:2405.21047}   ,
  eprint = {2405.21047},
  primaryclass = {cs},
  publisher = {arXiv},
  doi = {10.48550/arXiv.2405.21047},
  url = {http://arxiv.org/abs/2405.21047},
}

@misc{saad2025senai,
  title = {{{SENAI}}: {{Towards Software Engineering Native Generative Artificial Intelligence}}},
  author = {Saad, Mootez and L{\'o}pez, Jos{\'e} Antonio Hern{\'a}ndez and Chen, Boqi and Ernst, Neil and Varr{\'o}, D{\'a}niel and Sharma, Tushar},
  year = {2025},
  number = {arXiv:2503.15282},
  eprint = {2503.15282},
  primaryclass = {cs},
  publisher = {arXiv},
  doi = {10.48550/arXiv.2503.15282},
  url = {http://arxiv.org/abs/2503.15282},
}

@article{haque2025llms,
	title = {{LLMs: A game-changer for software engineers?}},
	volume = {5},
	issn = {27724859},
	shorttitle = {{LLMs}},
	url = {https://doi.org/10.1016/j.tbench.2025.100204}, 
	doi = {10.1016/j.tbench.2025.100204},
	journal = {BenchCouncil Trans. Benchmarks Stand.},
	author = {Haque, Md. Asraful},
	year = {2025},
	pages = {100204},
}

@misc{wang2024oop,
  title = {{{OOP}}: {{Object-Oriented Programming Evaluation Benchmark}} for {{Large Language Models}}},
  author = {Wang, Shuai and Ding, Liang and Shen, Li and Luo, Yong and Du, Bo and Tao, Dacheng},
  year = {2024},
  number = {arXiv:2401.06628}   ,
  eprint = {2401.06628},
  primaryclass = {cs},
  publisher = {arXiv},
  doi = {10.48550/arXiv.2401.06628},
  url = {http://arxiv.org/abs/2401.06628},
}

@misc{han2025graphrag,
  title = {Retrieval-{{Augmented Generation}} with {{Graphs}} ({{GraphRAG}})},
  author = {Han, Haoyu and others},
  year = {2025},
  number = {arXiv:2501.00309},
  eprint = {2501.00309},
  primaryclass = {cs},
  publisher = {arXiv},
  doi = {10.48550/arXiv.2501.00309},
  url = {http://arxiv.org/abs/2501.00309},
}

@misc{hou2025mcp,
  title = {{Model Context Protocol (MCP): Landscape, Security Threats, and Future Research Directions}}, 
  author = {Xinyi Hou and Yanjie Zhao and Shenao Wang and Haoyu Wang},
  year = {2025},
  eprint = {2503.23278},
  primaryClass = {cs.CR},
  url={https://arxiv.org/abs/2503.23278}, 
}

@online{ltm2mini,
  title = {{LTM-2-mini AI technology page}},
  author = {Magic},
  year = 2025,
  url = {https://lablab.ai/tech/ltm-2-mini},
}

@misc{bar2025ai,
  title = {{{AI}} for {{Code Synthesis}}: {{Can LLMs Generate Secure Code}}?},
  author = {Bar, Kaushik},
  year = {2025},
  publisher = {Elsevier BV},
  doi = {10.2139/ssrn.5157837},
  url = {https://dx.doi.org/10.2139/ssrn.5157837}, 
}

@misc{bhatt2023purple,
  title = {Purple {{Llama CyberSecEval}}: {{A Secure Coding Benchmark}} for {{Language Models}}},
  author = {Bhatt, Manish and others},
  year = {2023},
  number = {arXiv:2312.04724},
  eprint = {2312.04724},
  primaryclass = {cs},
  publisher = {arXiv},
  doi = {10.48550/arXiv.2312.04724},
  url = {http://arxiv.org/abs/2312.04724},
}

@inproceedings{thakur2025judging,
	title = {Judging the {Judges}: {Evaluating} {Alignment} and {Vulnerabilities} in {LLMs}-as-{Judges}},
	url = {https://aclanthology.org/2025.gem-1.33/},
	booktitle = {{GEM}²},
	author = {Thakur, Aman Singh and others},
	year = {2025},
	pages = {404--430},
}

@misc{jain2024livecodebench,
  title = {{{LiveCodeBench}}: {{Holistic}} and {{Contamination Free Evaluation}} of {{Large Language Models}} for {{Code}}},
  author = {Jain, Naman and others},
  year = {2024},
  number = {arXiv:2403.07974}  ,
  eprint = {2403.07974},
  primaryclass = {cs},
  publisher = {arXiv},
  doi = {10.48550/arXiv.2403.07974},
  url = {http://arxiv.org/abs/2403.07974},
  urldate = {2025-07-17},
  archiveprefix = {arXiv}
}

@inproceedings{yang2022natural,
  title = {{Natural Attack for Pre-Trained Models of Code}},
  booktitle = {ICSE '22},
  author = {Yang, Zhou and Shi, Jieke and He, Junda and Lo, David},
  year = {2022},
  pages = {1482--1493},
  publisher = {ACM},
  doi = {10.1145/3510003.3510146},
  url = {https://doi.org/10.1145/3510003.3510146}    
}

@misc{yang2024seccodeplt,
  title = {{{SecCodePLT}}: {{A Unified Platform}} for {{Evaluating}} the {{Security}} of {{Code GenAI}}},
  author = {Yang, Yu and others},
  year = {2024},
  number = {arXiv:2410.11096},
  eprint = {2410.11096},
  primaryclass = {cs},
  publisher = {arXiv},
  doi = {10.48550/arXiv.2410.11096},
  url = {http://arxiv.org/abs/2410.11096},
}

@misc{awal2025large,
  title = {Large {{Language Models}} as {{Robust Data Generators}} in {{Software Analytics}}: {{Are We There Yet}}?},
  author = {Awal, Md Abdul and Rochan, Mrigank and Roy, Chanchal K.},
  year = 2025,
  number = {arXiv:2411.10565}   ,
  eprint = {2411.10565},
  primaryclass = {cs},
  publisher = {arXiv},
  doi = {10.48550/arXiv.2411.10565},
  url = {http://arxiv.org/abs/2411.10565},
}

@misc{atemkeng2024ethics,
  title = {Ethics of {{Software Programming}} with {{Generative AI}}: {{Is Programming}} without {{Generative AI}} Always Radical?},
  author = {Atemkeng, Marcellin and others},
  year = {2024},
  number = {arXiv:2408.10554}   ,
  eprint = {2408.10554},
  primaryclass = {cs},
  publisher = {arXiv},
  doi = {10.48550/arXiv.2408.10554},
  url = {http://arxiv.org/abs/2408.10554},
}

@article{huang2025bias,
	title = {Bias {Testing} and {Mitigation} in {LLM}-based {Code} {Generation}},
	issn = {1049-331X, 1557-7392},
	url = {https://doi.org/10.1145/3724117}    ,
	doi = {10.1145/3724117},
	journal = {ACM TOSEM},
	author = {Huang, Dong and Zhang, Jie M. and Bu, Qingwen and Xie, Xiaofei and Chen, Junjie and Cui, Heming},
	year = {2025},
	pages = {3724117},
}

@misc{twist2025study,
  title = {A {{Study}} of {{LLMs}}' {{Preferences}} for {{Libraries}} and {{Programming Languages}}},
  author = {Twist, Lukas and Zhang, Jie M. and Harman, Mark and Syme, Don and Noppen, Joost and others},
  year = {2025},
  number = {arXiv:2503.17181}    ,
  eprint = {2503.17181},
  primaryclass = {cs},
  publisher = {arXiv},
  doi = {10.48550/arXiv.2503.17181 },
  url = {http://arxiv.org/abs/2503.17181},
}

@inproceedings{luccioni2024environmental,
  author    = {Sasha Luccioni and
               Bruna Trevelin and
               Margaret Mitchell},
  title     = {{The Environmental Impacts of AI -- Policy Primer}},
  booktitle = {Hugging Face Blog},
  year      = {2024},
  url       = {https://doi.org/10.57967/hf/3004}    ,
  doi       = {10.57967/hf/3004 }
}

@inproceedings{faiz2024llmcarbon,
  title = {{LLMCarbon: Modeling the End-to-End Carbon Footprint of Large Language Models}},
  author = {Ahmad Faiz and others},
  booktitle = {ICLR 2024},
  year = {2024},
  url = {https://openreview.net/forum?id=aIok3ZD9to}
}

@online{greensoftware,
  author    = {Hussain, Asim},
  title     = {{Green Software Foundation}},
  year      = {2025},
  url       = {https://greensoftware.foundation},
  publisher = {Joint Development Foundation Projects, LLC}
}

@inproceedings{cheung2025comparative,
  title = {{Comparative Analysis of Carbon Footprint in Manual vs. LLM-Assisted Code Development}},
  booktitle = {ResponsibleSE '25},
  author = {Cheung, Kuen Sum and Kaul, Mayuri and Jahangirova, Gunel and Mousavi, Mohammad Reza and Zie, Eric},
  year = 2025,
  pages = {13--20},
  publisher = {ACM},
  doi = {10.1145/3711919.3728678},
  url = {https://doi.org/10.1145/3711919.3728678} ,
  isbn = {979-8-4007-1461-0}
}

@inproceedings{cihan2024automated,
author = {Cihan, Umut and Haratian, Vahid and Icoz, Arda and Gul, Mert Kaan and Devran, Omercan and Bayendur, Emircan Furkan and Ucar, Baykal Mehmet and Tüzün, Eray},
booktitle = {ICSE-SEIP},
title = {{Automated Code Review in Practice}},
year = {2025},
volume = {},
ISSN = {},
pages = {425-436},
doi = {10.1109/ICSE-SEIP66354.2025.00043},
url = {https://doi.org/10.1109/ICSE-SEIP66354.2025.00043}  ,
publisher = {IEEE},
}

@techreport{dora2024,
  title	= {{DORA Accelerate State of DevOps 2024 Report}},
  author	= {Derek DeBellis and Kevin Storer and Amanda Lewis and others},
  year	= {2024},
  URL	= {https://dora.dev/research/2024/dora-report/},
  institution	= {Google}
}

@techreport{dora2025,
  title	= {{DORA Impact of Generative AI in Software Development}},
  author	= {Derek DeBellis and Kevin Storer and Daniella Villalba and others},
  year	= {2025},
  URL	= {https://dora.dev/research/ai/gen-ai-report/},
  institution	= {Google}
}

@techreport{gitclear2025ai,
  title = {{{GitClear AI Copilot Code Quality}}},
  author = {Harding, William},
  year = {2025},
  institution = {Alloy.dev Research},
  url = {https://www.gitclear.com/ai_assistant_code_quality_2025_research}
}

@misc{campos2024things,
  title = {{Some Things Never Change: How Far Generative {{AI}} Can Really Change Software Engineering Practice}},
  author = {de Campos, Aline and others},
  year = {2024},
  number = {arXiv:2406.09725}          ,
  eprint = {2406.09725},
  primaryclass = {cs},
  publisher = {arXiv},
  doi = {10.48550/arXiv.2406.09725},
  url = {http://arxiv.org/abs/2406.09725},
}

@misc{mastropaolo2024rise,
  title = {The {{Rise}} and {{Fall}}(?) Of {{Software Engineering}}},
  author = {Mastropaolo, Antonio and {Escobar-Vel{\'a}squez}, Camilo and {Linares-V{\'a}squez}, Mario},
  year = {2024},
  number = {arXiv:2406.10141}  ,
  eprint = {2406.10141},
  primaryclass = {cs},
  publisher = {arXiv},
  doi = {10.48550/arXiv.2406.10141},
  url = {http://arxiv.org/abs/2406.10141},
}

@online{stackoverflow2024survey,
  title = {2024 {{Developer Survey}}},
  author = {{Stack Overflow}},
  year = {2024},
  url = {https://survey.stackoverflow.co/2024/},
  urldate = {2025-07-25}
}

@inproceedings{beheshti2024natural,
	title = {Natural {Language}-{Oriented} {Programming} ({NLOP}): {Towards} {Democratizing} {Software} {Creation}},
	isbn = {979-8-3503-6851-2},
	url = {https://doi.org/10.1109/SSE62657.2024.00047}  ,
	doi = {10.1109/SSE62657.2024.00047},
	booktitle = {IEEE SSE 2024},
	author = {Beheshti, Amin},
	year = {2024},
	pages = {258--267},
}

@misc{hassan2024ainative,
  title = {Towards {{AI-Native Software Engineering}} ({{SE}} 3.0): {{A Vision}} and a {{Challenge Roadmap}}},
  author = {Hassan, Ahmed E. and others},
  year = {2024},
  number = {arXiv:2410.06107} ,
  eprint = {2410.06107},
  primaryclass = {cs},
  publisher = {arXiv},
  doi = {10.48550/arXiv.2410.06107},
  url = {http://arxiv.org/abs/2410.06107},
}

@article{qiu2025todays,
	title = {From {Today}’s {Code} to {Tomorrow}’s {Symphony}: {The} {AI} {Transformation} of {Developer}’s {Routine} by 2030},
	volume = {34},
	issn = {1049-331X, 1557-7392},
	url = {https://doi.org/10.1145/3709353} ,
	doi = {10.1145/3709353},
	language = {en},
	number = {5},
	journal = {ACM TOSEM},
	author = {Qiu, Ketai and Puccinelli, Niccolò and Ciniselli, Matteo and Di Grazia, Luca},
	year = {2025},
	pages = {1--17},
}

@misc{jin2025advancing,
  title = {Towards {{Advancing Code Generation}} with {{Large Language Models}}: {{A Research Roadmap}}},
  author = {Jin, Haolin and Chen, Huaming and Lu, Qinghua and Zhu, Liming},
  year = {2025},
  number = {arXiv:2501.11354}  ,
  eprint = {2501.11354},
  primaryclass = {cs},
  publisher = {arXiv},
  doi = {10.48550/arXiv.2501.11354},
  url = {http://arxiv.org/abs/2501.11354},
}

@misc{jin2024mare,
  title = {{{MARE}}: {{Multi-Agents Collaboration Framework}} for {{Requirements Engineering}}},
  author = {Jin, Dongming and Jin, Zhi and Chen, Xiaohong and Wang, Chunhui},
  year = {2024},
  number = {arXiv:2405.03256}  ,
  eprint = {2405.03256},
  primaryclass = {cs},
  publisher = {arXiv},
  doi = {10.48550/arXiv.2405.03256},
  url = {http://arxiv.org/abs/2405.03256},
}

@misc{fowler2024panel,
  title = {Panel {{Discussion}}: {{Where Is Software Development Going}}?},
  author = {{Fowler, Martin}},
  year = {2024},
  url = {https://gotocph.com/2024/sessions/3332/panel-discussion-where-is-software-development-going},
}

@article{terragni2025future,
  title = {The {{Future}} of {{AI-Driven Software Engineering}}},
  author = {Terragni, Valerio and Vella, Annie and Roop, Partha and Blincoe, Kelly},
  year = {2025},
  journal = {ACM TOSEM},
  issn = {1049-331X},
  doi = {10.1145/3715003},
  url = {https://doi.org/10.1145/3715003}  ,
}

@inproceedings{alami2025human,
	title = {Human and {Machine}: {How} {Software} {Engineers} {Perceive} and {Engage} with {AI}-{Assisted} {Code} {Reviews} {Compared} to {Their} {Peers}},
	isbn = {979-8-3315-3871-2},
	shorttitle = {Human and {Machine}},
	url = {https://doi.org/10.1109/CHASE66643.2025.00016}  ,
	doi = {10.1109/CHASE66643.2025.00016 },
	booktitle = {CHASE 2025},
	author = {Alami, Adam and Ernst, Neil},
	year = {2025},
	pages = {63--74},
}

@article{salehie2009self,
author = {Salehie, Mazeiar and Tahvildari, Ladan},
title = {Self-adaptive software: Landscape and research challenges},
year = {2009},
publisher = {Association for Computing Machinery},
volume = {4},
number = {2},
issn = {1556-4665},
url = {https://doi.org/10.1145/1516533.1516538}   ,
doi = {10.1145/1516533.1516538},
journal = {ACM Trans. Auton. Adapt. Syst.},
}

@techreport{gartner2024ai,
  type = {White {{Paper}}},
  title = {{{AI Will Not Replace Software Engineers}} (and {{May}}, in {{Fact}}, {{Require More}})},
  author = {Philip Walsh and others},
  year = {2024},
  institution = {Gartner},
  url = {https://www.gartner.com/en/documents/5724051},
}

@article{he2025llmbased,
	title = {{LLM}-{Based} {Multi}-{Agent} {Systems} for {Software} {Engineering}: {Literature} {Review}, {Vision}, and the {Road} {Ahead}},
	volume = {34},
	issn = {1049-331X, 1557-7392},
	shorttitle = {{LLM}-{Based} {Multi}-{Agent} {Systems} for {Software} {Engineering}},
	url = {https://doi.org/10.1145/3712003}   ,
	doi = {10.1145/3712003},
	number = {5},
	journal = {ACM TOSEM},
	author = {He, Junda and Treude, Christoph and Lo, David},
	year = {2025},
	pages = {1--30},
}

@misc{hao2024training,
  title = {Training {{Large Language Models}} to {{Reason}} in a {{Continuous Latent Space}}},
  author = {Hao, Shibo and others},
  year = {2024},
  number = {arXiv:2412.06769}   ,
  eprint = {2412.06769},
  primaryclass = {cs},
  publisher = {arXiv},
  doi = {10.48550/arXiv.2412.06769},
  url = {http://arxiv.org/abs/2412.06769},
}

@misc{lindsey2025biology,
	title = {On the {Biology} of a {Large} {Language} {Model}},
	note = {{Anthropic}},
	url = {https://transformer-circuits.pub/2025/attribution-graphs/biology.html},
	author = {Lindsey, Jack and Gurnee, Wes and Ameisen, Emmanuel},
	year = {2025},
}

@misc{euaiact2025transparency, 
  title = {{Key Issue 5: Transparency Obligations - EU AI Act}}, 
  url = {https://www.euaiact.com/key-issue/5}, 
  journal = {EU AI Act - EU Artificial Intelligence Act}, 
  author = {{E}uropean Union}, 
  year = {2025}
}

@online{litan2024tackling,
  title = {Tackling {{Trust}}, {{Risk}} and {{Security}} in {{AI Models}}},
  author = {Litan, Avivah},
  year = 2024,
  url = {https://www.gartner.com/en/articles/ai-trust-and-ai-risk}
}

@techreport{polemi2023multilayer,
  title = {{{Multilayer Framework for Good Cybersecurity Practices for AI}}},
  author = {Polemi, Nineta and Pra{\c c}a, Isabel},
  year = 2023,
  institution = {ENISA},
  url = {https://www.enisa.europa.eu/publications/multilayer-framework-for-good-cybersecurity-practices-for-ai}
}

@misc{euaiact2024,
  author       = {{European Data Protection Supervisor}},
  title        = {{AI Act Regulation (EU) 2024/1689}},
  publisher    = {Publications Office of the European Union},
  year         = {2024},
  url          = {https://data.europa.eu/doi/10.2804/4225375}     ,
}

@misc{voskanyan2025codexsentinel,
  title = {{{CodexSentinel}} - {{Advanced Go Security}} \& {{Code Analysis Platform}}},
  author = {Voskanyan, Voskan},
  year = 2025,
  url = {https://github.com/Voskan/codexsentinel},
  urldate = {2025-10-28}
}

@inproceedings{zhang2024seccoder,
	title = {{SecCoder}: {Towards} {Generalizable} and {Robust} {Secure} {Code} {Generation}},
	url = {https://aclanthology.org/2024.emnlp-main.806/},
	doi = {10.18653/v1/2024.emnlp-main.806},
	booktitle = {EMNLP 2024},
	publisher = {ACL},
	author = {Zhang, Boyu and others},
	year = {2024},
	pages = {14557--14571},
}

@incollection{naik2024comparative,
	title = {A {Comparative} {Analysis} of {Threat} {Modelling} {Methods}: {STRIDE}, {DREAD}, {VAST}, {PASTA}, {OCTAVE}, and {LINDDUN}},
	volume = {884},
    url = {https://doi.org/10.1007/978-3-031-74443-3_16}   ,
	booktitle = {C3AI 2024},
	publisher = {Springer},
	author = {Naik, Nitin and others},
	year = {2024},
	doi = {10.1007/978-3-031-74443-3_16},
	pages = {271--280},
}

@online{owasp2023mlsec,
  author       = {{OWASP Foundation}},
  title        = {{OWASP Machine Learning Security Top Ten}},
  year         = {2023},
  url          = {https://owasp.org/www-project-machine-learning-security-top-10/}
}

@misc{codecarbon2025,
  author       = {Benoit Courty and others},
  title        = {mlco2/codecarbon: v2.4.1},
  year         = 2024,
  publisher    = {Zenodo},
  version      = {v2.4.1},
  doi          = {10.5281/zenodo.11171501},
  url          = {https://doi.org/10.5281/zenodo.11171501}    , 
}

@techreport{isoiec30165,
  author = {ISO/IEC},
  title     = {{Internet of Things (IoT) — Real-time IoT framework}},
  institution = {IEC Central Office},
  number    = {TR 30165},
  year      = {2021},
  url       = {https://www.iso.org/standard/53285.html}
}

@online{greendigital2023,
  author       = {{International Telecommunication Union}},
  title        = {{Green Digital Action}},
  year         = {2023},
  url          = {https://www.itu.int/initiatives/green-digital-action/},
}

@online{chatgpt,
  year             = {2025},
  author           = {{OpenAI Inc.}},
  url              = {https://chatgpt.com/},
  title            = {{ChatGPT}}
}

@online{cursor,
  year             = 2025,
  author           = {{Anysphere Inc.}},
  url              = {https://cursor.com/},
  title            = {{Cursor}}
}

@online{google2025q3,
  year             = {2025},
  author           = {Google},
  title            = {Q3 earnings call: CEO’s remarks},
  url              = {https://blog.google/inside-google/message-ceo/alphabet-earnings-q3-2024/}
}

@inproceedings{tankelevitch2024metacognitive,
  title = {The {{Metacognitive Demands}} and {{Opportunities}} of {{Generative AI}}},
  booktitle = {CHI '24},
  author = {Tankelevitch, Lev and others},
  year = {2024},
  eprint = {2312.10893},
  primaryclass = {cs},
  pages = {1--24},
  doi = {10.1145/3613904.3642902 },
  url = {https://doi.org/10.1145/3613904.3642902}   ,
  urldate = {2025-03-01},
  archiveprefix = {arXiv}
}

@misc{kwa2025measuring,
  title = {Measuring {{AI Ability}} to {{Complete Long Tasks}}},
  author = {Kwa, Thomas and others},
  year = {2025},
  number = {arXiv:2503.14499  }  ,
  eprint = {2503.14499},
  primaryclass = {cs},
  publisher = {arXiv},
  doi = {10.48550/arXiv.2503.14499 },
  url = {http://arxiv.org/abs/2503.14499},
}

@online{saifi2025let,
  author       = {Sohail Saifi},
  title        = {{I Let ChatGPT Make All My Architectural Decisions for a Month: The Surprising Results}},
  year         = {2025},
  day          = {28},
  organization    = {Medium},
  url          = {https://medium.com/@sohail_saifi/i-let-chatgpt-make-all-my-architectural-decisions-for-a-month-the-surprising-results-e21f1254c74c},
}

@online{wikipedia2025vibecoding,
  author       = {{Wikipedia contributors}},
  title        = {Vibe coding},
  year         = {2025},
  organization    = {Wikipedia, The Free Encyclopedia},
  url          = {https://en.wikipedia.org/wiki/Vibe_coding},
  note         = {Accessed: 2025-10-29}
}

@online{langchain2024,
  author       = {{LangChain AI, Inc.}},
  title        = {{LangChain}},
  url          = {https://www.langchain.com/},
  year         = {2024},
}

@article{mike2022computational,
  title = {{Computational Thinking in the Era of Data Science}},
  author = {Mike, Koby and Ragonis, Noa and {Rosenberg-Kima}, Rinat B. and Hazzan, Orit},
  year = 2022,
  journal = {Communications of the ACM},
  volume = {65},
  number = {8},
  pages = {33--35},
  issn = {0001-0782, 1557-7317},
  doi = {10.1145/3545109 },
  url = {https://doi.org/10.1145/3545109}  ,
}

@online{siemensdev,
  author    = {{Siemens AG}},
  title     = {{Siemens Developer Portal}},
  year      = {2025},
  url       = {https://developer.siemens.com},
  note      = {Accessed: 2025-10-29},
  publisher = {Siemens}
}

\end{document}